\documentclass{aa}  

\usepackage{graphicx}
\usepackage{txfonts}
\usepackage{natbib}
\usepackage{multirow}

\begin{document}

\title{Observation of CH$_{3}$$^{17}$OH and CH$_{3}$$^{18}$OH in Orion KL: A New Tool to Study Star-Formation History}
\titlerunning{CH$_3$$^{17}$OH and CH$_3$$^{18}$OH in Oropn KL}

\author{
Yoshimasa Watanabe\inst{1,2},
{Takahiro Oyama}\inst{2},
{Akemi Tamanai}\inst{2,3},
{Shaoshan Zeng}\inst{2},
{Nami Sakai}\inst{2}
}
\authorrunning{Watanabe et al.}

\institute{
Shibaura Institute of Technology, 3-7-5 Toyosu, Koto-ku, Tokyo 135-8548, Japan; \email{nabe@shibaura-it.ac.jp}
\and RIKEN Pioneering Research Institute, 2-1, Hirosawa, Wako, Saitama 351-0198, Japan
\and Max Planck Institute for Astronomy (MPIA), K\"{o}nigstuhl 17, D-69117 Heidelberg, Germany
}

\date{Received July 9, 2025; accepted October 9, 2025}
\abstract
{}
{
Methanol 
is a seed species of complex organic molecules that is of fundamental importance in astrochemistry.  Although various isotopologues of CH$_3$OH have been detected in the interstellar medium (ISM), CH$_{3}$$^{17}$OH is only tentatively detected in Sgr~B2.  To confirm the presence of CH$_{3}$$^{17}$OH in the ISM and to investigate its abundance, we search for its emission lines in the Orion~KL region. 
}
{
We have obtained image cubes covering the frequency ranges 236.40~GHz-236.65~GHz and 231.68~GHz-231.88~GHz with a resolution of $\sim 2$ arcsec using ALMA archival data observed toward the Orion~KL region.  The spectra detected at the two CH$_3$$^{18}$OH peaks, MeOH1 and MeOH2, are compared to the spectrum and frequencies of CH$_{3}$$^{17}$OH measured in the laboratory.  The column densities of CH$_3$$^{17}$OH and CH$_3$$^{18}$OH are estimated under the assumption of local thermodynamic equilibrium condition with fixed excitation temperatures. 
}
{
We have identified six emission lines of CH$_{3}$$^{17}$OH in MeOH1 and MeOH2 and confirmed that the line profiles and spatial distributions are consistent with those of CH$_3$$^{18}$OH. The abundance ratios of CH$_3$$^{18}$OH/CH$_3$$^{17}$OH are evaluated to be $\sim 3.4-3.5$ and are similar to the canonical value of $^{18}$O/$^{17}$O $\sim 3-4$ derived from CO observations in the Orion~KL region.  We have compared the results with the previous study of CH$_3$OH and evaluated CH$_3$$^{16}$OH/CH$_3$$^{17}$OH ratios to be $\sim 2300-2500$ at a resolution of $\sim 4$~arcsec.  The ratios are close to the $^{16}$O/$^{17}$O ratio in the local ISM.    
}
{
This result indicates that the CH$_3$OH isotopologues can serve as new tracers of oxygen isotope ratios in star-forming regions because the opacity of CH$_3$OH can be evaluated using transition lines spanning a wide range of line intensities.  Moreover, this method enables us to study the star-formation history of our Galaxy with the aid of the Galactic chemical evolution models.
}

\maketitle

\begin{figure*}[t]
\centering
\includegraphics[clip,width=16cm,bb = 0 0 537.985 264.941]{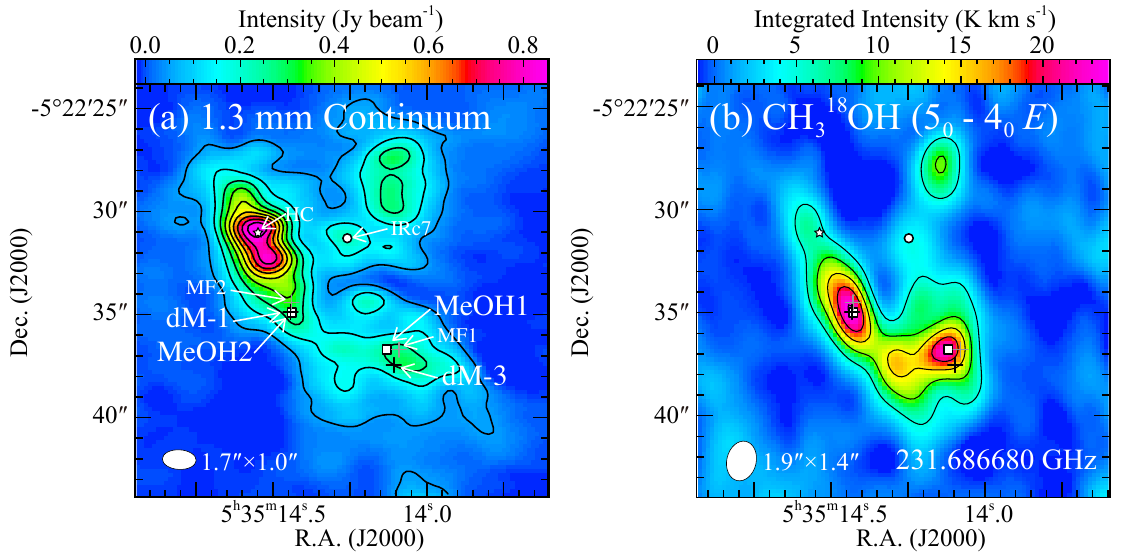}
\caption{(a) 1.3 mm continuum image of Orion KL observed with ALMA (\#2013.1.00553.S).  Contour levels are from 50~mJy~beam$^{-1}$ ($5\sigma$) to 950~mJy~beam$^{-1}$ with a 100~mJy~beam$^{-1}$ step. (b) Integrated intensity image of CH$_3$$^{18}$OH ($5_0-4_0$ $E$) obtained by the ALMA SV.  Contour levels are from 4.8 K~km~s$^{-1}$ ($3\sigma$) to 28.8 K~km~s$^{-1}$ with a 4.8 K~km~s$^{-1}$ step. White ellipses at the left bottom corner 
 indicate the synthesized beams.  The CH$_3$$^{18}$OH peaks (MeOH1 and MeOH2) are shown by the white squares.  The positions of Hot Core (HC), IRc7, and the HCOOCH$_3$ peaks (MF) identified by \citet{Favre2011} are shown by the star, circle, and gray cross marks, respectively.  The black cross marks show the positions of dM-1 and dM-3.}
\label{fig1}
\end{figure*}

\section{Introduction}\label{sec1:intro}

Methanol (CH$_3$OH) is one of the most fundamental saturated organic molecules in interstellar medium (ISM) and a seed species of complex organic molecules.  A laboratory experiment has demonstrated that this molecule is efficiently produced by the successive hydrogenation of CO on the surface of icy dust grains, as supported by theoretical and experimental studies \citep[e.g.,][]{Tielens1997,Watanabe2002}. In cold molecular clouds, the CH$_3$OH formed is thought to be released into the gas phase through non-thermal sublimation driven by energy, such as surplus energy in chemical reactions, although the efficiency is not very high \citep[e.g.,][]{Garrod2007, Soma2015}. 
On the other hand, the CH$_3$OH in the inner envelope of protostellar core appears in the gas phase through desorption by protostellar heating.  Such regions are called hot cores in massive-star forming regions, or hot corinos in low-mass star forming regions \citep[e.g.][]{Blake1987,Dishoeck1995,Ceccarelli2007,Yang2021}.  CH$_3$OH has also been observed toward various objects including shocked regions induced by outflows from protostars \citep[e.g.][]{Bachiller1997}, 
and external galaxies \citep[e.g.,][]{Henkel1987,Watanabe2014}.  

The $^{18}$O isotopolog of CH$_3$OH has been observed in various objects \citep[e.g.][]{Gardner1989,Gelder2022}, while the observation of the $^{17}$O isotopolog has only been reported in a single case \citep{Muller2024}.  The primary reason for this was the lack of available laboratory spectroscopic data until recently.  However, experimental spectroscopic measurements of CH$_3$$^{17}$OH have now been carried out and reported by \citet[][]{Muller2024} and \citet[][]{Tamanai2025}.   \citet{Muller2024} has presented rotational line list of CH$_3$$^{17}$OH and the tentative detection of it in Sgr B2(N2b).  \citet{Tamanai2025} also conducted spectroscopic measurement from 216~GHz to 264~GHz by an emission-type millimeter and submillimeter spectrometer \citep{Watanabe2021} and reported line parameters of CH$_3$$^{17}$OH based on the spectroscopic data and previous study \citep{Hoshino1996}.  Moreover, they assigned several transition lines of $^{13}$CH$_3^{17}$OH, which were contained in the CH$_3$$^{17}$OH sample as the natural abundance of $^{13}$C.  The study of CH$_3$$^{17}$OH in star-forming regions is still in its infancy, and the abundance ratios between CH$_3$$^{17}$OH and other CH$_3$OH isotopologues containing different oxygen isotopes are poorly understood.  In contrast, deuterated CH$_3$OH isotopologues are known to be enhanced by 2-3 orders of magnitude compared with the cosmic H/D ratio of $\sim10^{-5}$ \citep[e.g.][]{Mauersberger1988,Peng2012}, due to grain-surface and gas-phase chemical reactions \citep[e.g.][]{Charnley1997}.  Meanwhile, the abundance of $^{13}$CH$_3$OH is known to be similar to the elemental abundance ratio of $^{12}$C/$^{13}$C \citep{Soma2015}.  This is because CH$_3$OH molecules are mainly formed from CO molecules, which account for the majority of the carbon budget in the molecular clouds.  In this study, we search for the CH$_3$$^{17}$OH emission lines in the Orion~KL region by using the ALMA archive data and evaluate the isotope ratios of $^{18}$O/$^{17}$O, $^{16}$O/$^{17}$O, and $^{16}$O/$^{18}$O in CH$_3$OH.


Orion~KL is nearby massive star forming region at the distance of $\sim 420$~pc \citep[e.g.,][]{Hirota2007,Menten2007}.  In this region, oxygen-bearing molecular species, such as CH$_3$OH and HCOOCH$_3$ are preferentially distributed in the vicinity of the compact ridge, while nitrogen-bearing molecular species, such as CH$_3$CN and CH$_3$CH$_2$CN, are associated around the hot core \citep[e.g.,][]{Wright1996,2005Beuther,Tercero2018}.  
\cite{Favre2011} identified 28 methyl formate emission peaks and found that the strongest peaks, MF1 and MF2, are in the compact ridge.  \cite{Peng2012} also identified the strongest CH$_2$DOH emission peaks dM-1 and dM-3 in the almost identical positions as MF2 and MF1, respectively. As previous studies have demonstrated that CH$_3$OH shows a high column density and thus a high abundance in the compact ridge, we investigated the presence of CH$_3$$^{17}$OH lines in this region by using Atacama Large Millimeter/submillimeter Array (ALMA). 

\section{Data and Reductions}
We utilized two ALMA archival data ADS/JAO.ALMA \#2013.1.00553.S (PI: P. Goldsmith) and ADS/JAO.ALMA \#2011.0.00009.SV (the ALMA Science Verification data of Orion KL) to search for CH$_3$$^{17}$OH and CH$_3$$^{18}$OH, respectively.  The observation of ADS/JAO.ALMA \#2013.1.00553.S was conducted on 29 December 2014 toward the Orion KL region where the phase center was (R.A.(J2000), Dec.(J2000)) = ($05^{\rm h}35^{\rm m}14^{\rm s}.160$, $-05^{\circ}22'08''.504$).  The details regarding the observation are described in \cite{Pagani2017}.  We used one spectral window that covered the frequency range from 236.26~GHz to 237.20~GHz with a frequency resolution of 488~kHz. The calibrated visibility data were acquired by applying the data reduction script provided by the ALMA observatory with the Common Astronomy Software Application (CASA) package \citep{casa2022} version 4.2.2 which contained the pipeline for the Cycle 2 ALMA observation.  The visibility was imaged by using the CLEAN algorithm using Briggs weighting with a robustness parameter of 0.5, after the continuum emission was subtracted from the visibility data by fitting a 0$^{\rm th}$-order polynomial to the line free channels using task \textit{uvcontsub}.  The sensitivity and the synthesized beam of the cube data are 7~mJy~beam$^{-1}$ and $1''.7 \times 1''.0$ with PA $=87.6^{\circ}$, respectively.  The continuum image was made by the continuum visibility data generated by the task \textit{uvcontsub}.  Figure~\ref{fig1} (a) shows the continuum image in the 1.3~mm band (236.7~GHz).  The image reproduces the structures reported by \citet{Pagani2017} which used the same observation data.  

The observation of ADS/JAO.ALMA \#2011.0.00009.SV was carried out on 20 January 2012 at the phase center of (R.A.(J2000), Dec.(J2000)) = ($05^{\rm h}35^{\rm m}14^{\rm s}.350$, $-05^{\circ}22'36''.350$) as a part of the ALMA Science Verification to demonstrate the capabilities of spectral line survey observation with ALMA.  The observation covered the frequency range from 214~GHz to 247~GHz with a frequency resolution of 488~kHz.  Among the frequency range, we obtained synthesized image from 231.68~GHz to 231.88~GHz, witch contained 16 transitions of the $R$-branch series of CH$_3$$^{18}$OH with $J=5-4$.  The calibrated visibility data was imaged by using the CLEAN algorithm by using Briggs weighting with a robustness parameter of 0.5 after the continuum emission was subtracted from the visibility data by fitting the 0$^{\rm th}$-order polynomial to the line free channels.  The sensitivity and the synthesized beam of cube data are 12~mJy~beam$^{-1}$ and $1''.9 \times 1''.4$ with PA $=-12.8^{\circ}$, respectively.  

All the cube data and continuum images were corrected for the primary beam attenuation by applying task \textit{impbcor}.  CASA version 5.6.3 was used for all data processing after the calibration process. 

\begin{figure*}[t!]
\centering
\includegraphics[clip,width=16cm,bb = 0.000 0.000 795.962 915.521]{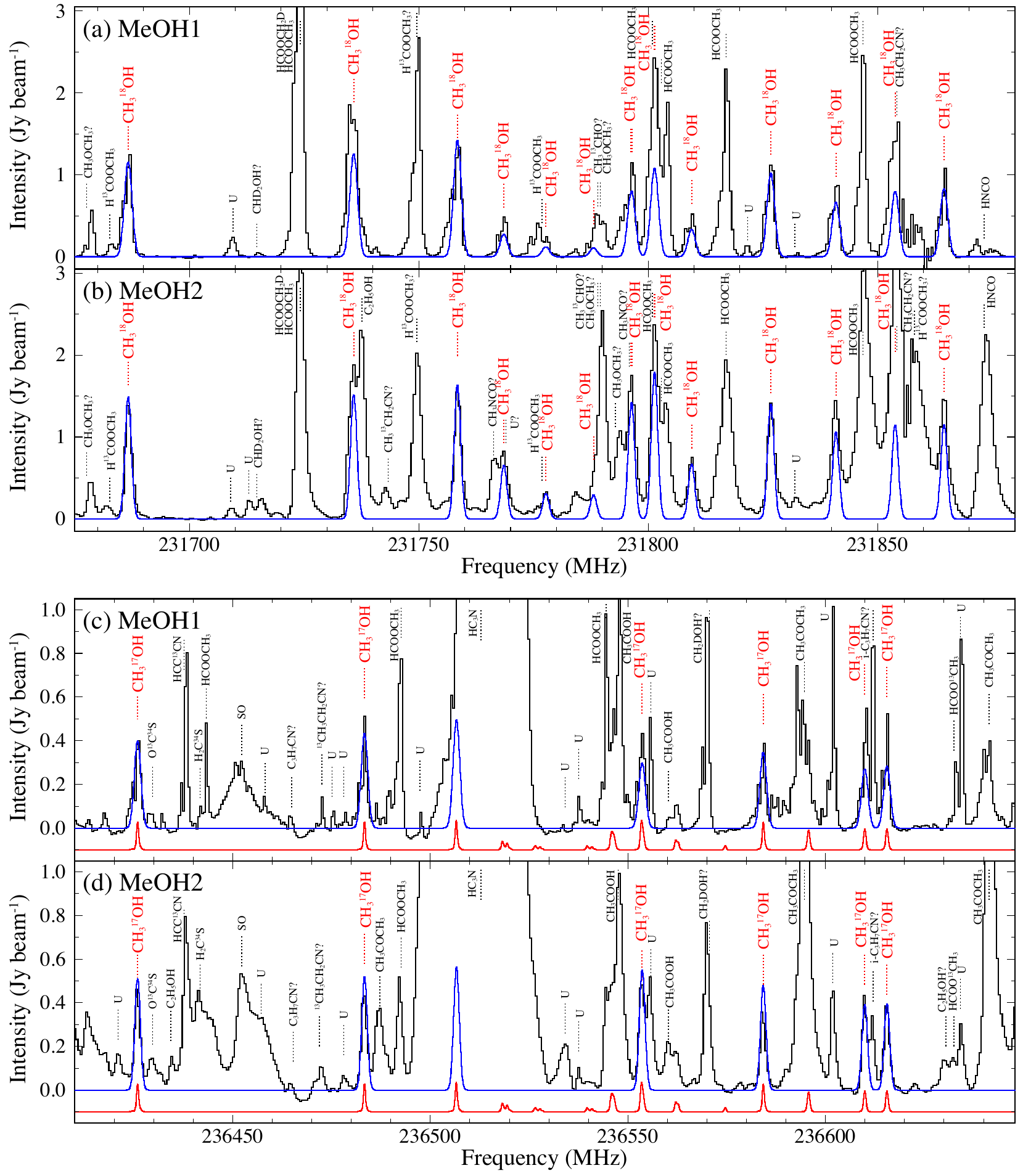}
\caption{CH$_3$$^{18}$OH spectra (black line) observed at the CH$_3$$^{18}$OH peaks (a) MeOH1 and (b) MeOH2 in the rest frame, and CH$_3$$^{17}$OH spectra (black line) observed at (c) MeOH1 and (d) MeOH2.  The system velocities are assumed to be $V_{\rm LSR} = 7.50$ km~s$^{-1}$ and 7.55 km~s$^{-1}$ for MeOH1 and MeOH2, respectively.  The spectra are obtained after the angular resolutions of observation data are convolved to be the $2''.0 \times 2''.0$ beam.  The solid red line in (c) and (d) is the experimentally measured CH$_3$$^{17}$OH spectrum obtained in a laboratory \citep{Tamanai2025} with an arbitrary intensity scale.  Red dashed lines are the positions of CH$_3$$^{18}$OH and CH$_3$$^{17}$OH and black dashed lines are those of other molecules.  Molecular names with a question mark "?" indicate the potential molecular species.  "U" indicates an unidentified emission line.  Blue lines are model spectra of CH$_3$$^{18}$OH and CH$_3$$^{17}$OH predicted with the LTE model by using the column densities estimated in Table~\ref{tab3} and the line parameters listed in Table~\ref{tab2}.  The line width is assumed to be 2.5~km~s$^{-1}$ and 2.1~km~s$^{-1}$ for MeOH1 and MeOH2, respectively, which are evaluated from the Gaussian fitting to the CH$_3$$^{18}$OH ($5_0-4_0,\,E$) line.  The $S\mu^2$ and $E_{\rm u}$ values are taken from \citet{Tamanai2025}.}
\label{fig2}
\end{figure*}

\begin{figure*}[t!]
\centering
\includegraphics[clip,width=15.5cm,bb = 0 0 654.016 405.616]{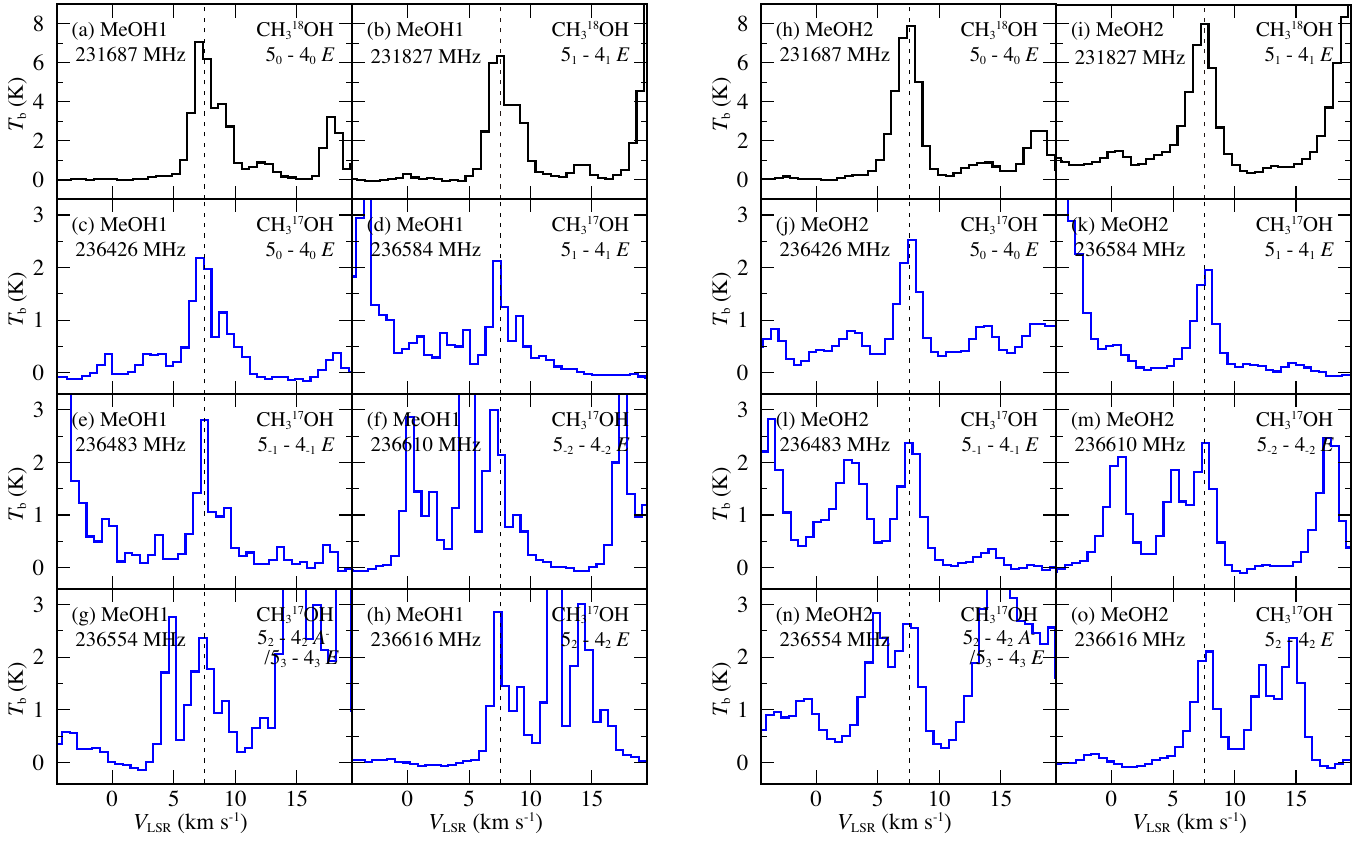}
\caption{Line profiles of CH$_3$$^{18}$OH and CH$_3$$^{17}$OH (Table~\ref{tab1}) at the two methanol peaks MeOH1 (left: a - h) and MeOH2 (right: i - o).  The spectra are obtained after the angular resolutions of observation data are convolved to be the $2''.0 \times 2''.0$ beam for fair comparisons.  The vertical scale is the brightness temperature.  The vertical dashed lines indicate the system velocities of $V_{\rm LSR} = 7.50$ km~s$^{-1}$ and 7.55 km~s$^{-1}$ for MeOH1 and MeOH2, respectively. }
\label{fig3}
\end{figure*}

\section{Results}
\subsection{Detections of CH$_3$$^{18}$OH and CH$_3$$^{17}$OH}

In the frequency range from 231.68~GHz to 231.88~GHz (Figure~\ref{fig2} (a) and (b)), we detected the emission lines of CH$_3$$^{18}$OH ($5_0-4_0$ $E$) with a signal-to-noise (SN) ratio of more than $100\sigma$.  Figure~\ref{fig1} (b) is the integrated intensity map of CH$_3$$^{18}$OH ($5_0-4_0$ $E$) which is made by integrating the velocity range from 4.5~km~s$^{-1}$ to 11.5~km~s$^{-1}$.  We found two prominent CH$_3$$^{18}$OH peaks, MeOH1 and MeOH2 (Table~\ref{tab1}), which coincide with the methyl formate peaks, MF1 and MF2 \citep{Favre2011}, respectively, within the synthesized beam of $1''.9 \times 1''.4$.  In addition to the CH$_3$$^{18}$OH ($5_0-4_0$ $E$), 12 emission lines of CH$_3$$^{18}$OH are detected at the MeOH1 and MeOH2 in the frequency range, but most of them are contaminated by other molecules except for CH$_3$$^{18}$OH ($5_0-4_0$ $A^+$) and CH$_3$$^{18}$OH ($5_1-4_1$ $E$) (Figure~\ref{fig2} (a) and (b)). 

\begin{table}
\centering
\caption{Positions of MeOH1 and MeOH2 \label{tab1}}
\begin{tabular}{cccc}
\hline\hline
Name & R.A. (J2000) & Dec. (J2000) & $V_{\rm LSR}$ \tablefootmark{a}\\
 & &  & (km~s$^{-1}$)  \\
\hline
MeOH1 & $05^{\rm h} 35^{\rm m} 14^{\rm s}.13$ &  $-05^{\circ} 22' 36''.7$ & 7.50 \\
MeOH2 & $05^{\rm h} 35^{\rm m} 14^{\rm s}.44$ &  $-05^{\circ} 22' 34''.9$ & 7.55 \\
\hline
\end{tabular}
\tablefoot{
\tablefoottext{a}{The systemic velocity at the local standard of rest (LSR).  The velocity is estimated by the Gaussian fit to the spectrum of CH$_3$$^{18}$OH ($5_0-4_0$ $E$).}
}
\end{table}
\begin{table}
\centering
\caption{Line Parameters of Identified CH$_3$$^{18}$OH and CH$_3$$^{17}$OH \label{tab2}}
\begin{tabular}{cccrr}
\hline\hline
Molecule & Transition & Freq. & $E_{\rm u}$ \tablefootmark{a} & S$\mu^2$ \\
 & & (GHz) & (K) & (Debye$^2$) \\
\hline
CH$_3$$^{18}$OH \tablefootmark{b} & $5_{ 0}-4_{ 0}$ $E$   & 231.686680 &  46.22 & 16.166 \\ 
CH$_3$$^{18}$OH \tablefootmark{b} & $5_{-1}-4_{-1}$ $E$   & 231.735830 &  38.99 & 15.569 \\ 
CH$_3$$^{18}$OH \tablefootmark{b} & $5_{ 0}-4_{ 0}$ $A^+$ & 231.758446 &  33.38 & 16.194 \\ 
CH$_3$$^{18}$OH \tablefootmark{b} & $5_{ 4}-4_{ 4}$ $A^-$ & 231.768541 & 114.26 &  5.825 \\ 
CH$_3$$^{18}$OH \tablefootmark{b} & $5_{ 4}-4_{ 4}$ $A^+$ & 231.768542 & 114.26 &  5.825 \\ 
CH$_3$$^{18}$OH \tablefootmark{b} & $5_{-4}-4_{-4}$ $E$   & 231.777708 & 121.82 &  5.829 \\ 
CH$_3$$^{18}$OH \tablefootmark{b} & $5_{ 4}-4_{ 4}$ $E$   & 231.788110 & 129.55 &  5.867 \\ 
CH$_3$$^{18}$OH \tablefootmark{b} & $5_{ 3}-4_{ 3}$ $A^+$ & 231.796218 &  83.49 & 10.336 \\ 
CH$_3$$^{18}$OH \tablefootmark{b} & $5_{ 3}-4_{ 3}$ $A^-$ & 231.796521 &  83.49 & 10.336 \\ 
CH$_3$$^{18}$OH \tablefootmark{b} & $5_{ 3}-4_{ 3}$ $E$   & 231.801304 &  81.31 & 10.372 \\ 
CH$_3$$^{18}$OH \tablefootmark{b} & $5_{ 2}-4_{ 2}$ $A^-$ & 231.801466 &  70.86 & 13.755 \\ 
CH$_3$$^{18}$OH \tablefootmark{b} & $5_{-3}-4_{-3}$ $E$   & 231.809480 &  96.02 & 10.450 \\ 
CH$_3$$^{18}$OH \tablefootmark{b} & $5_{ 1}-4_{ 1}$ $E$   & 231.826744 &  54.14 & 16.171 \\ 
CH$_3$$^{18}$OH \tablefootmark{b} & $5_{ 2}-4_{ 2}$ $A^+$ & 231.840925 &  70.87 & 13.756 \\ 
CH$_3$$^{18}$OH \tablefootmark{b} & $5_{-2}-4_{-2}$ $E$   & 231.853853 &  59.26 & 13.625 \\ 
CH$_3$$^{18}$OH \tablefootmark{b} & $5_{ 2}-4_{ 2}$ $E$   & 231.864501 &  55.77 & 13.380 \\ 
CH$_3$$^{17}$OH \tablefootmark{c} & $5_{0}-4_{0}$ $E$     & 236.426222 &  47.09 &  4.041 \\
CH$_3$$^{17}$OH \tablefootmark{c} & $5_{-1}-4_{-1}$ $E$   & 236.483332 &  39.68 &  3.887 \\
CH$_3$$^{17}$OH \tablefootmark{c} & $5_{ 2}-4_{ 2}$ $A^-$ & 236.553557 &  71.71 & 13.380 \\ 
CH$_3$$^{17}$OH \tablefootmark{c} & $5_{3}-4_{3}$ $E$     & 236.553786 &  82.09 &  2.590 \\
CH$_3$$^{17}$OH \tablefootmark{c} & $5_{1}-4_{1}$ $E$     & 236.584222 &  55.04 &  4.020 \\
CH$_3$$^{17}$OH \tablefootmark{c} & $5_{-2}-4_{-2}$ $E$   & 236.609861 &  60.04 &  3.403 \\
CH$_3$$^{17}$OH \tablefootmark{c} & $5_{2}-4_{2}$ $E$     & 236.615575 &  56.47 &  3.345 \\
\hline 
\end{tabular}
\tablefoot{
\tablefoottext{a}{The upper state energy.}
\tablefoottext{b}{The values are taken from CDMS \citep{Muller2001} and are based on spectroscopic measurements by \citet{Hughes1951}, \citet{Gerry1976}, \citet{Hoshino1996}, \citet{Predoi1997}, \citet{Ikeda1998}, and \citet{Fisher2007} .}
\tablefoottext{c}{The values are taken from \citep{Tamanai2025}. The effect from quadrupole hyperfine structure arising as a result of the nuclear spin of $^{17}$O (I = 5/2) is within the range of error in frequency.}
}
\end{table}

\begin{table*}[t!]
\centering
\caption{Column densities \label{tab3}}
\begin{tabular}{ccccccc}
\hline\hline
Position & beam & Molecule & Column Density & ${\rm O}^{18}/{\rm O}^{17}$ & ${\rm O}^{16}/{\rm O}^{17}$ & ${\rm O}^{16}/{\rm O}^{18}$ \\
 & & & (cm$^{-2}$) & & & \\
\hline
\multirow{2}{*}{MeOH1} & \multirow{2}{*}{$2''.0 \times 2''.0$} & CH$_3$$^{18}$OH \tablefootmark{a,c}& $(3.7 \pm 0.2)\times 10^{15}$ & \multirow{2}{*}{$3.3 \pm 0.2$} & \multirow{2}{*}{N.A.} & \multirow{2}{*}{N.A.} \\
 & & CH$_3$$^{17}$OH \tablefootmark{a,d}& $(1.1 \pm 0.6)\times 10^{15}$ & & & \\
\hline
\multirow{2}{*}{MeOH2} & \multirow{2}{*}{$2''.0 \times 2''.0$} & CH$_3$$^{18}$OH \tablefootmark{b,c} & $(9.8 \pm 0.5)\times 10^{15}$ &  \multirow{2}{*}{$3.3 \pm 0.2$} & \multirow{2}{*}{N.A.} & \multirow{2}{*}{N.A.}\\
 & & CH$_3$$^{17}$OH \tablefootmark{b,d} & $(3.0 \pm 0.2)\times 10^{15}$ & & & \\
\hline
\multirow{3}{*}{dM-1 \tablefootmark{e}} & \multirow{3}{*}{$3''.8 \times 2''.0$} & CH$_3$$^{18}$OH \tablefootmark{b,c} & $(6.5 \pm 0.3)\times 10^{15}$ &  \multirow{3}{*}{$3.5 \pm 0.3$} & \multirow{3}{*}{$2300 \pm 300$} & \multirow{3}{*}{$650 \pm 70$}\\
 & & CH$_3$$^{17}$OH \tablefootmark{b,d} & $(1.9 \pm 0.1)\times 10^{15}$ & & & \\
 & & CH$_3$OH \tablefootmark{b,f} & $(4.2 \pm 0.4)\times 10^{18}$ & & & \\
\hline
\multirow{3}{*}{dM-3 \tablefootmark{g}} & \multirow{3}{*}{$3''.8 \times 2''.0$} & CH$_3$$^{18}$OH \tablefootmark{a,c} & $(2.4 \pm 0.1)\times 10^{15}$ &  \multirow{3}{*}{$3.4 \pm 0.3$} & \multirow{3}{*}{$2500 \pm 200$} & \multirow{3}{*}{$710 \pm 60$}\\
 & & CH$_3$$^{17}$OH \tablefootmark{a,d} & $(6.9 \pm 0.4)\times 10^{14}$ & & & \\
 & & CH$_3$OH \tablefootmark{a,f} & $(1.7 \pm 0.1)\times 10^{18}$ & & & \\
\hline
\end{tabular}
\tablefoot{
\tablefoottext{a}{The rotation temperature of 58~K derived by \citet{Peng2012} is assumed.}
\tablefoottext{b}{The rotation temperature of 130~K derived by \citet{Peng2012} is assumed.}
\tablefoottext{c}{The column density is derived from a single transition line of CH$_3$$^{18}$OH ($5_0 - 4_0$ $E$).}
\tablefoottext{d}{The column density is derived from a single transition line of CH$_3$$^{17}$OH ($5_0 - 4_0$ $E$).}
\tablefoottext{e}{The position of dM-1 is $(\alpha_{\rm J2000}, \delta_{\rm J2000})=(05^{\rm h} 35^{\rm m} 14^{\rm s}.442,-05^{\circ} 22' 34''.86)$.}
\tablefoottext{f}{The column density is estimated by \citet{Peng2012}.}
\tablefoottext{g}{The position of dM-3 is $(\alpha_{\rm J2000}, \delta_{\rm J2000})=(05^{\rm h} 35^{\rm m} 14^{\rm s}.107, -05^{\circ} 22' 37''.43)$.}
}
\end{table*}

\begin{figure*}[t!]
\centering
\includegraphics[clip,width=15.5cm,bb = 0 0 882.053 541.403]{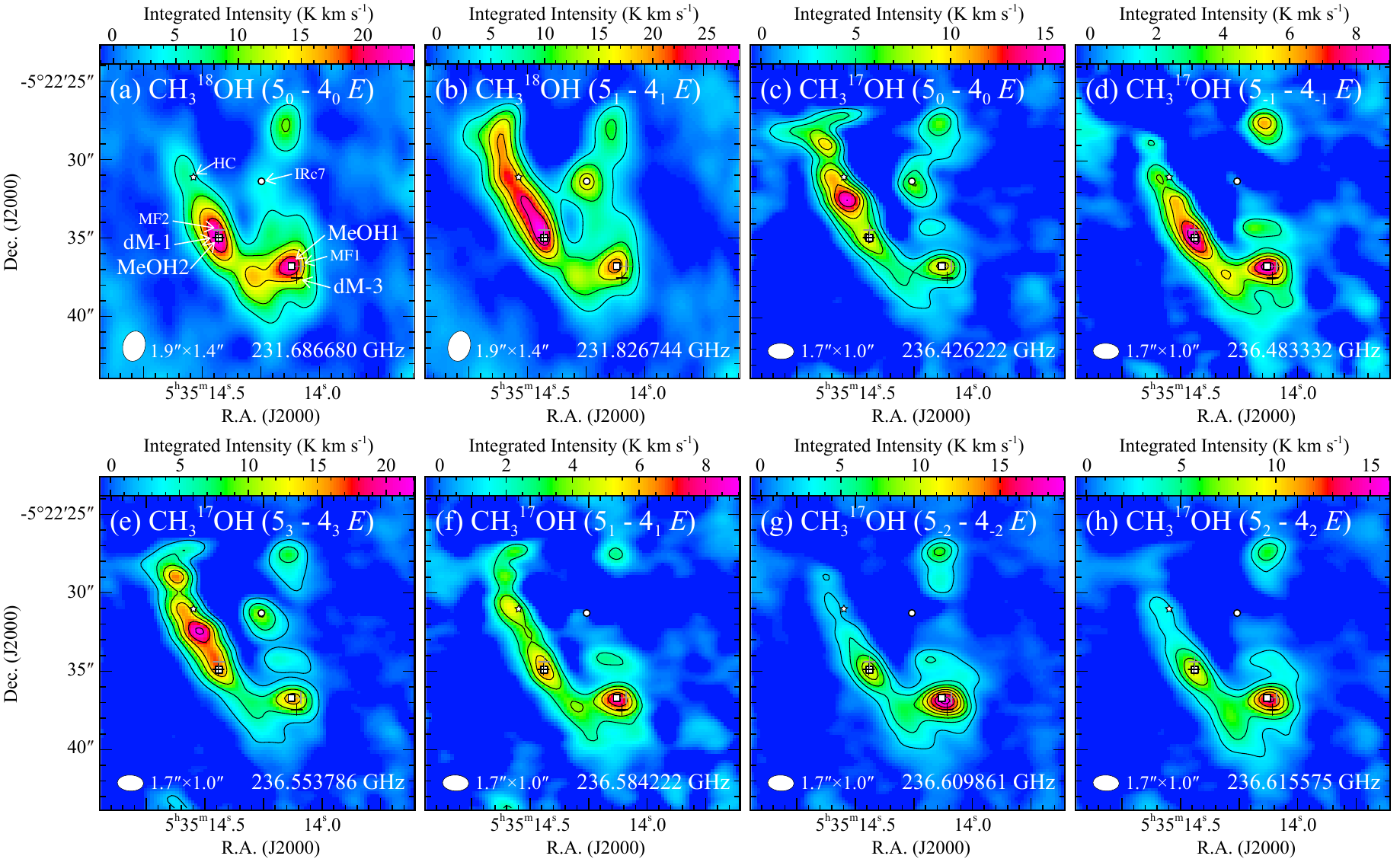}
\caption{Integrated intensity maps of (a) CH$_3$$^{18}$OH ($5_0 - 4_0$ $E$), (b) CH$_3$$^{18}$OH ($5_1 - 4_1$ $E$), (c) CH$_3$$^{17}$OH ($5_0 - 4_0$ $E$), (d) CH$_3$$^{17}$OH ($5_{-1} - 4_{-1}$ $E$), (e) CH$_3$$^{17}$OH ($5_3 - 4_3$ $E$), (f) CH$_3$$^{17}$OH ($5_1 - 4_1$ $E$), (g) CH$_3$$^{17}$OH ($5_{-2} - 4_{-2}$ $E$), and (h) CH$_3$$^{17}$OH ($5_2 - 4_2$ $E$).  The contour levels are (a) from 4.8 K~km~s$^{-1}$ ($3\sigma$) to 19.2 K~km~s$^{-1}$ with a 4.8 K~km~s$^{-1}$ step, (b) from 4.8 K~km~s$^{-1}$ ($3\sigma$) to 24.0 K~km~s$^{-1}$ with a 4.8 K~km~s$^{-1}$ step, (c) from 2.3 K~km~s$^{-1}$ ($3\sigma$) to 11.3 K~km~s$^{-1}$ with a 2.3 K~km~s$^{-1}$ step, (d) from 2.3 K~km~s$^{-1}$ ($3\sigma$) to 8.4 K~km~s$^{-1}$ with a 1.5 K~km~s$^{-1}$ step, (e) from 2.3 K~km~s$^{-1}$ ($3\sigma$) to 21.3 K~km~s$^{-1}$ with a 3.8 K~km~s$^{-1}$ step, (f) from 4.8 K~km~s$^{-1}$ ($3\sigma$) to 8.4 K~km~s$^{-1}$ with a 1.5 K~km~s$^{-1}$ step, (g) from 4.8 K~km~s$^{-1}$ ($3\sigma$) to 18.2 K~km~s$^{-1}$ with a 2.3 K~km~s$^{-1}$ step, and (h) from 2.3 K~km~s$^{-1}$ ($3\sigma$) to 13.6 K~km~s$^{-1}$ with a 2.3 K~km~s$^{-1}$ step.  White ellipses at the left bottom corner indicate the synthesized beams.  The symbols shown in the maps are the same as those in Figure~\ref{fig1}.}
\label{fig4}
\end{figure*}

We identified CH$_3$$^{17}$OH emission lines in the spectra from 236.26~GHz to 237.20~GHz at the two peaks (Figure~\ref{fig2} (c) and (d)).  The red spectrum shows the CH$_3$$^{17}$OH spectrum obtained from spectroscopic measurements in our laboratory \citep{Tamanai2025}, for comparison with the observed spectra (black line).  In MeOH1 and MeOH2, six emission lines are identified as CH$_3$$^{17}$OH with a S/N ratio of more than $50\sigma$.  Most of the detected spectral lines are $E$ symmetry transitions, whereas a $A$ symmetry transition of $5_2-4_2~A^-$ is found to be blended with the $5_3-4_3~E$ line at 236.554~GHz.  In addition to the CH$_3$$^{17}$OH lines, we assigned the other molecular lines in the spectra with the aid of spectral line databases, the Cologne Database for Molecular Spectroscopy \citep[CDMS:][]{Muller2001} and the Submillimeter, Millimeter, and Microwave Spectral Line Catalog provided by Jet Propulsion Laboratory \citep{Pickett1998}.  Figure~\ref{fig3} shows the line profiles of CH$_3$$^{17}$OH and CH$_3$$^{18}$OH.  Here, the cube data was convolved with Gaussian kernels to compare the spectra at the same angular resolution of $2''.0 \times 2''.0$.  The peak velocities and line profiles of CH$_3$$^{17}$OH lines are similar to those of CH$_3$$^{18}$OH lines, although some of the CH$_3$$^{17}$OH lines are contaminated by other nearby molecular emission lines.  The line profile shows two velocity components centered at $\sim 7.5$~km~s$^{-1}$ and $\sim 9$~km~s$^{-1}$ in MeOH1.  The same velocity components have been detected by the observation of methyl formate in MF1 \citep{Favre2011}.  The integrated intensity maps of CH$_3$$^{17}$OH and CH$_3$$^{18}$OH lines are shown in Figure~\ref{fig4}.  The maps were obtained by integrating the velocity range from 4.5~km~s$^{-1}$ to 11.5~km~s$^{-1}$.  Most of CH$_3$$^{17}$OH images show peak emission at the two CH$_3$$^{18}$OH peaks MeOH1 and MeOH2, while some images (e.g., Figure~\ref{fig4} (b) and (d)) show the strongest peaks at other positions due to contaminations of other molecular emission lines.

\subsection{Column densities}

The column densities of CH$_3$$^{17}$OH and CH$_3$$^{18}$OH were calculated in MeOH1 and MeOH2 under the assumption of local thermodynamic equilibrium (LTE) approximation with optically thin conditions (Table~\ref{tab3}).  Since we are observing transition lines with the critical density of $n_{\rm crit} \sim 10^5$~cm$^{-3}$ \footnote{The $n_{\rm crit}$ value is estimated from $A_{i,i}/C_{i,j}$ for CH$_3$OH ($5_0-4_0$, $E$) at 100~K where $A_{i,j}$ and $C_{i,j}$ is the Einstein A and C coefficient, respectively, from Leiden Atomic and Molecular Database \citep{Schoier2005}.}, the assumption of LTE is justified because the emissions originate from sufficiently dense regions with the H$_2$ volume density of $n_{\rm H_2} > 10^8$~cm$^{-3}$ \citep{Favre2011}.  We used the integrated intensities of CH$_3$$^{18}$OH ($5_0-4_0$ $E$) and CH$_3$$^{17}$OH ($5_0-4_0$ $E$) obtained from the convolved data cubes which have the common angular resolutions of $2''.0 \times 2''.0$.  Using the CDMS and JPL databases, we have confirmed that these emission line are less contaminated by emission lines of other interstellar molecules.  Assuming the canonical values of $^{16}$O/$^{18}$O $=560$ and $^{16}$O/$^{17}$O $=2000$ \citep{Wilson1999} and using the column densities and rotational temperatures of CH$_3$OH at dM-1 and dM-3 from \citet{Peng2012}, the optical depths of CH$_3$$^{18}$OH ($5_0-4_0$ $E$) and CH$_3$$^{17}$OH ($5_0-4_0$ $E$) are expected to be $0.05-0.12$ and $0.01-0.04$, respectively.  Therefore, the assumption of the optically thin condition is considered reasonable.  In the calculation, we assumed the rotation temperature of 58~K and 130~K for MeOH1 and MeOH2, respectively, which are the rotation temperatures of CH$_3$OH in dM-3 and dM-1 \citep[][]{Peng2012}.   Since we evaluated the column densities from the single transitions with the fixed rotation temperatures, the column densities are subject to uncertainties due to these factors.    

By using the assumed rotation temperatures and the derived column densities of CH$_3$$^{17}$OH and CH$_3$$^{18}$OH, we generated model spectra under the LTE approximation and overlaid them on the observed spectra at MeOH1 and MeOH2 (blue spectra in Figure~\ref{fig2}).  Here, we utilized the line parameters of CH$_3$$^{17}$OH by \citet{Tamanai2025} and those in CDMS for CH$_3$$^{18}$OH (Table~\ref{tab2}).  The model spectra show good agreement with the observed spectra at the positions of CH$_3$$^{17}$OH and CH$_3$$^{18}$OH, except for the lines contaminated by emission lines from other molecules.  For the CH$_3$$^{18}$OH, we also calculated the column densities using the integrated intensities of less contaminated CH$_3$$^{18}$OH ($5_1-4_1$ $E$) line in order to confirm the values.  These are consistent, within the uncertainties, with the values obtained from CH$_3$$^{18}$OH ($5_0-4_0$ $E$).  In this analysis, we attempted to estimate column density and rotation temperature by using the rotation diagram method, however, the limited range of the upper state energies ($E_{\rm u} = 40-60$~K) resulted in significant uncertainties in the derived values.  Therefore, we employed the rotation temperatures of CH$_3$OH from \citet{Peng2012}.

The CH$_3$$^{18}$OH/CH$_3$$^{17}$OH ratios are evaluated to be $3.3\pm 0.2$ in both MeOH1 and MeOH2, respectively.  The assumed rotation temperatures have little effect on the CH$_3$$^{18}$OH/CH$_3$$^{17}$OH ratios because the upper state energies are very similar for the two transition lines due to the same quantum numbers.  

In addition to the column densities in MeOH1 and MeOH2, the column densities of the two CH$_3$OH isotopologues were calculated in the two CH$_2$DOH peaks, dM-1 and dM-3, because \citet{Peng2012} estimated the column densities of CH$_3$OH main isotopologue by using the observation in the 101~GHz band with IRAM PdBI.  In spite of the high column density of CH$_3$OH in the regions, the optical depths of the CH$_3$OH lines have been confirmed to be less than 0.2 \citep{Peng2012} due to the relatively small $S\mu^2$ values of these transition lines.  The positions of dM-1 and dM-3 almost coincide with those of MeOH2 and MeOH1, respectively, within the beam size of the PdBI observation (Figure~\ref{fig1}).  We used the same procedures to calculate the column densities of the two CH$_3$OH isotopologues, after smoothing the angular resolution of the ALMA data to be $3''.8 \times 2''.0$ with PA of $22^{\circ}$ which is the observation beam of \citet{Peng2012}.  The rotation temperatures of 130~K and 58~K were adopted for dM-1 and dM-3, respectively, following \citet{Peng2012}.  The column densities of three isotopologues are summarized in Table~\ref{tab3}.  The evaluated CH$_3$$^{16}$OH/CH$_3$$^{17}$OH, CH$_3$$^{16}$OH/CH$_3$$^{18}$OH, and CH$_3$$^{18}$OH/CH$_3$$^{17}$OH ratios are $2300\pm 300$, $650\pm 70$, and $3.5\pm 0.3$, respectively in dM-1.

\section{Discussion}
\subsection{Comparisons of oxygen isotope ratios with those of other molecules}
The isotopic ratios of $^{18}$O/$^{17}$O in the interstellar matter have been estimated from the observations of CO isotopologues \citep[e.g.,][]{Penzias1981}.  \citet{Persson2007} evaluated the $^{18}$O/$^{17}$O ratio of $3.6 \pm 0.7$ by using C$^{18}$O($J=5-4$) and C$^{17}$O($J=5-4$) observed with Odin satellite toward Orion KL.  \citet{Plume2012} derived the $^{18}$O/$^{17}$O ratio of $4.1^{+2.1}_{-1.3}$ in the Compact Ridge, in which MeOH1 is associated, by using high-$J$ lines of CO isotopologues observed with Herschel/HIFI while they estimated different values of the ratios for the Hot Core ($3.0^{+1.2}_{-1.1}$), Outflow/Plateau ($1.7^{+0.4}_{-0.5}$), and Extended Ridge ($2.3\pm 0.5$).  \citet{Zou2023} estimated the $^{18}$O/$^{17}$O ratio of $4.1\pm0.1$ in G209.00-19.38 in Orion Nebula by using C$^{18}$O($J=2-1$) and C$^{17}$O($J=2-1$) lines observed with SMT 10~m telescope.  In addition, many transition lines of H$_2^{18}$O and H$_2$$^{17}$O have been detected toward Orion KL with Herschel/HIFI \citep{Melnick2010,Neill2013}, but the $^{18}$O/$^{17}$O ratio was not estimated from the water isotopologues because the detected H$_2$$^{18}$O emission lines are moderately optically thick \citep{Neill2013}.

The $^{18}$O/$^{17}$O ratios derived from CH$_3$OH isotopologues are consistent with the ratios derived from high-$J$ lines of CO isotopologues within the uncertainties.  On the other hand, the ratio derived from the CO isotopologue lines with low-$J$ \citep{Zou2023} is significantly higher than those from CH$_3$OH isotopologues.  The discrepancy cannot be explained by the optical depth of low-$J$ transition lines of CO isotopologues.  The $^{18}$O/$^{17}$O ratio is likely to be underestimated, since the optical depth of C$^{18}$O is usually higher than that of C$^{17}$O by a factor of several.  One possible explanation is that C$^{17}$O is dissociated by the selective photodissociation by the interstellar UV radiation because C$^{17}$O is optically thinner than C$^{18}$O as discussed by \citet{Plume2012}.  As a result, the $^{18}$O/$^{17}$O ratio derived from CO isotopologues would be higher than the canonical ratio.  The transition lines of CO with low-$J$ especially tend to trace extended diffuse molecular gas which can be affected by the interstellar UV radiation.  On the other hand, CH$_3$OH is thought to reside in the deep inside of dense gas and to be less affected by UV radiation.  The other potential explanation is that the CH$_3$$^{18}$OH line is moderately optically thick. 

The $^{16}$O/$^{17}$O ratio of $2300 \pm 300$ and $^{16}$O/$^{18}$O ratios of $650 \pm 70$ in dM-1 are consistent with the ratios of $^{16}$O/$^{17}$O $=2000\pm 200$ and $^{16}$O/$^{18}$O $=560\pm 30$ in the local ISM \citep{Wilson1999} within the uncertainty.  In contrast, the $^{16}$O/$^{18}$O ratios are significantly higher than the isotopic ratios of methyl formate reported by \citet{Tercero2012}, which estimated the $^{16}$O/$^{18}$O ratio to be $\sim 218$ from the total column density of the two isotopomer HC$^{18}$OOCH$_3$ and HCO$^{18}$OCH$_3$ observed with IRAM~30~m telescope toward Orion~KL \citep{Tercero2010}.  The $^{16}$O/$^{18}$O ratio of CH$_3$OH should be compared with the ratio of HCOOCH$_3$/HC$^{18}$OOCH$_3$ or HCOOCH$_3$/HCO$^{18}$OCH$_3$ because HCOOCH$_3$ has two inequivalent oxygen atoms.  The probability that each oxygen atom is substituted by $^{18}$O in HCOOCH$_3$ is expected to be proportional to the $^{16}$O/$^{18}$O ratio if neither fractionation nor dilution process is occurred by chemical reactions.  Therefore, the $^{16}$O/$^{18}$O ratio in HCOOCH$_3$ is evaluated to be twice higher than the ratio by \citet{Tercero2012}.  Even if the factor of two is considered, our ratios are still higher than that of HCOOCH$_3$ by a factor of 1.5.  One of the reasons for the underestimation is that the column density of the main HCOOCH$_3$ isotopologue is underestimated due to the use of optically thick transitions.  Another possibility is the effect of missing flux caused by the different antenna configurations between the observations or the subthermal excitation condition of CH$_3$OH in this position as discussed in \citet{Peng2012}.  To confirm the difference, we need to observe multiple transition lines of CH$_3$OH and CH$_3$$^{18}$OH simultaneously with ALMA and apply an excitation analysis.  This confirmation is for future work.

\subsection{CH$_3$OH as a tool of isotope ratio measurement}
Although our analysis has uncertainty in the $^{16}$O/$^{18}$O ratio, the isotope ratios derived from CH$_3$OH isotopologue lines can potentially be more reliable value than those from the low-$J$ CO isotopologue lines because the CO isotopologues are sometimes affected by the optical thickness as well as the selective photodissociation by the intense interstellar UV radiation as aforementioned.  However, it should be noted that CH$_3$OH is formed by hydrogenation of CO that has been frozen on interstellar dust grains in cold environments \citep[e.g.,][]{Watanabe2002,Soma2015}. This implies that CH$_3$OH may preserve the isotopic ratios of CO from the prestellar phase when the gas density is relatively low. While CH$_3$OH is likely less affected by the selective photodissociation than the CO molecules that have remained exposed to interstellar UV radiation until the present day, the possible influence of this formation history on the derived isotope ratios warrants further investigation.

Specifically, future studies should examine the spatial distribution of CH$_3$$^{17}$OH and CH$_3$$^{18}$OH in hot cores or hot corinos with disk-like structures. By comparing the distributions from the outer regions where interstellar UV photons are more likely to penetrate to the inner regions where UV photons from the protostar are dominant, it may be possible to evaluate the effect of selective photodissociation on oxygen isotope ratios.  Such studies, especially when discussed in connection with the known variation of $^{18}$O/$^{16}$O and $^{17}$O/$^{16}$O ratios in meteorites in the Solar system \citep[e.g.,][]{Clayton1993,Yurimoto2004}, may provide valuable insights into the chemical origins of the Solar system.

The advantage of using CH$_3$OH isotopologues lies more in the following point than those mentioned above: if an appropriate frequency setting is selected, multiple transition lines of CH$_3$OH isotopologues can be simultaneously observed within a single ALMA frequency setup.  As a result, precise isotope ratios are derived from the LTE modeling or the non-LTE modeling if the collision rates are available.  Moreover, the column density of the main CH$_3$OH isotopologue is derived if optically thin transition lines with small line intensity $S\mu^2$ are carefully selected.  As described in Section 4.1, the optical depth of spectral lines is a critical factor in obtaining reliable isotopic ratio estimates, especially for the main isotopologue.  In our analysis, we employed the $J=5-4$ a-type transitions of CH$_3$$^{17}$OH and CH$_3$$^{18}$OH to derive the column density.  However, the same transition lines of the main isotopologue could not be used due to high optical depth.  Instead of these lines, we utilized the column densities of the main isotopologue which were derived from optically thin lines with smaller $S\mu^2$ values by \citet{Peng2012}.   Taking these into consideration, this method allows for a reliable estimation of oxygen isotopic ratios using this method.  The elemental abundances are usually measured by the observation of atomic lines toward HII regions.  The abundances of isotopes, however, cannot be measured at the same method because the atomic lines of isotopes are contaminated by the main isotopic line due to the relatively broad linewidth ($\sim 10$~km~s$^{-1}$) in the HII regions.  The isotope ratios are generally determined by the observation of isotopologues.

These isotope ratios can be used to constrain the star formation history of the Galaxy by the Galactic `chemical evolution' (GCE) models.  The elemental abundances reflect the star formation history in the Galaxy because the metal enrichment proceeds with nucleosynthesis in various types of stars.  The negative abundance gradients for heavy elements along the Galactocentric distance have been known by the observations of atomic lines toward the Galactic HII regions \citep[e.g.,][]{Shaver1983,Arellano2020}.  The observed gradients of elemental abundances are thought to result from the inside-out star formation in the Galaxy \citep[e.g.,][]{Larson1976} and are explained by the GCE models \citep[e.g., references in][]{Romano2022}.  Because different isotopes are synthesized by different processes, the isotopic ratios of stable isotopes provide unique parameters to constrain the GCE models.  For instance, the $^{12}$C/$^{13}$C ratio increases with the Galactocentric distance \citep{Milam2005} because the primary element of $^{12}$C is produced faster than the secondary element of $^{13}$C.  In the case of the oxygen isotopes, $^{18}$O is primarily produced in the massive star while $^{17}$O is synthesized in the intermediate-mass stars with longer timescale \citep[e.g.,][]{Henkel1993}.  \citet{Martin2019} reported that the $^{18}$O/$^{17}$O ratio is enhanced in the central molecular zone of starburst galaxy NGC~253 suggesting the enrichment of $^{18}$O by the massive stars.  By the observation of CO isotopologues, a positive gradient of the $^{18}$O/$^{17}$O ratio has been reported \citep[e.g.][]{Wouterloot2008,Zhang2020,Zou2023} in the Galaxy.  Nevertheless, the sample size is small and scarce on the outer disk ($> 10$~kpc from the Galactic center).  Therefore, simultaneous survey observations of CH$_3$$^{17}$OH, CH$_3$$^{18}$OH, and CH$_3$$^{16}$OH toward the hot cores on the outer disk regions with ALMA would increase the sample size of $^{18}$O/$^{17}$O and $^{16}$O/$^{18}$O ratios and contribute to improving the GCE models in our Galaxy.

\section{Conclusions}

In this paper, the identification of emission lines of CH$_3$$^{17}$OH in the Orion~KL region has been carried out by making use of our own experimentally measured spectra in the frequency range from 236.40~GHz to 236.65~GHz.  We evaluated the abundance ratios of the CH$_3$OH isotopologues, CH$_3$$^{16}$OH, CH$_3$$^{17}$OH, and CH$_3$$^{18}$OH through comparative verification.  The results are summarized as follows:
\begin{enumerate}
\item We identified six emission lines of rare isotopologue CH$_3$$^{17}$OH toward the two CH$_3$$^{18}$OH emission peaks, MeOH1 and MeOH2, in Orion~KL by using ALMA Archive data in the 1.3~mm band.  Because the line profiles and spatial resolution of CH$_3$$^{17}$OH are similar to those of CH$_3$$^{18}$OH, these CH$_3$OH isotopologues are expected to reside in the same regions.
\item The isotopologue ratios of CH$_3$$^{18}$OH/CH$_3$$^{17}$OH are derived to be $\sim 3.4\pm0.2$ on a resolution of 2.0~arcsec under the LTE approximation with fixed excitation temperatures.  The ratios are consistent with the $^{18}$O/$^{17}$O ratios estimated from high excitation transition lines of CO isotopologues observed in the Orion~KL region.
\item We also estimated the CH$_3$$^{16}$OH/CH$_3$$^{17}$OH ratio to be $\sim 2300\pm200$ in dM-1 of Orion~KL by using CH$_3$OH column density reported by \citet{Peng2012} on a resolution of $\sim 4.0$~arcsec and found that the ratios are consistent with the canonical $^{16}$O/$^{17}$O ratio in the local ISM.
\item The simultaneous observation of CH$_3$OH isotopologues with ALMA would provide a unique method to estimate the isotope ratios of oxygen in the star-forming regions.  Since transition lines of CH$_3$OH have wide range of the line intensities, optically thin transition lines can be used even for the main isotopologue.  The measurements of the oxygen isotope ratios will contribute to constraining the GCE models and delineating the star formation history in the Galaxy. 
\end{enumerate}

\begin{acknowledgements}
This paper makes use of the ALMA data set ADS/JAO.ALMA\#2013.1.00553.S and ADS/JAO.ALMA\#2011.0.00009.SV.  ALMA is a partnership of the ESO (representing its member states), the NSF (USA) and NINS (Japan), together with the NRC (Canada) and the NSC and ASIAA (Taiwan), in cooperation with the Republic of Chile. The Joint ALMA Observatory is operated by the ESO, the AUI/NRAO and the NAOJ. The authors are grateful to the ALMA staff for their excellent support.  This study is supported by a Grant-in-Aid from the Ministry of Education, Culture, Sports, Science, and Technology of Japan (No.20H05845) and a pioneering project in RIKEN (Evolution of Matter in the Universe).  Y.W. acknowledges support from a Grant-in-Aid from the Ministry of Education, Culture, Sports, Science, and Technology of Japan (No.25K01042,24K00675,25H00676).
\end{acknowledgements}

%
%
\bibliographystyle{aa}
\bibliography{AA56325-25-arxiv}

\begin{thebibliography}{56}
\expandafter\ifx\csname natexlab\endcsname\relax\def\natexlab#1{#1}\fi

\bibitem[{{Arellano-C{\'o}rdova} {et~al.}(2020){Arellano-C{\'o}rdova},
  {Esteban}, {Garc{\'\i}a-Rojas}, \& {M{\'e}ndez-Delgado}}]{Arellano2020}
{Arellano-C{\'o}rdova}, K.~Z., {Esteban}, C., {Garc{\'\i}a-Rojas}, J., \&
  {M{\'e}ndez-Delgado}, J.~E. 2020, \mnras, 496, 1051

\bibitem[{{Bachiller} \& {P{\'e}rez Guti{\'e}rrez}(1997)}]{Bachiller1997}
{Bachiller}, R. \& {P{\'e}rez Guti{\'e}rrez}, M. 1997, \apjl, 487, L93

\bibitem[{{Beuther} {et~al.}(2005){Beuther}, {Zhang}, {Greenhill}, {Reid},
  {Wilner}, {Keto}, {Shinnaga}, {Ho}, {Moran}, {Liu}, \& {Chang}}]{2005Beuther}
{Beuther}, H., {Zhang}, Q., {Greenhill}, L.~J., {et~al.} 2005, \apj, 632, 355

\bibitem[{{Blake} {et~al.}(1987){Blake}, {Sutton}, {Masson}, \&
  {Phillips}}]{Blake1987}
{Blake}, G.~A., {Sutton}, E.~C., {Masson}, C.~R., \& {Phillips}, T.~G. 1987,
  \apj, 315, 621

\bibitem[{{CASA Team} {et~al.}(2022){CASA Team}, {Bean}, {Bhatnagar}, {Castro},
  {Donovan Meyer}, {Emonts}, {Garcia}, {Garwood}, {Golap}, {Gonzalez Villalba},
  {Harris}, {Hayashi}, {Hoskins}, {Hsieh}, {Jagannathan}, {Kawasaki},
  {Keimpema}, {Kettenis}, {Lopez}, {Marvil}, {Masters}, {McNichols},
  {Mehringer}, {Miel}, {Moellenbrock}, {Montesino}, {Nakazato}, {Ott}, {Petry},
  {Pokorny}, {Raba}, {Rau}, {Schiebel}, {Schweighart}, {Sekhar}, {Shimada},
  {Small}, {Steeb}, {Sugimoto}, {Suoranta}, {Tsutsumi}, {van Bemmel},
  {Verkouter}, {Wells}, {Xiong}, {Szomoru}, {Griffith}, {Glendenning}, \&
  {Kern}}]{casa2022}
{CASA Team}, {Bean}, B., {Bhatnagar}, S., {et~al.} 2022, \pasp, 134, 114501

\bibitem[{{Ceccarelli} {et~al.}(2007){Ceccarelli}, {Caselli}, {Herbst},
  {Tielens}, \& {Caux}}]{Ceccarelli2007}
{Ceccarelli}, C., {Caselli}, P., {Herbst}, E., {Tielens}, A.~G.~G.~M., \&
  {Caux}, E. 2007, in Protostars and Planets V, ed. B.~{Reipurth}, D.~{Jewitt},
  \& K.~{Keil}, 47

\bibitem[{{Charnley} {et~al.}(1997){Charnley}, {Tielens}, \&
  {Rodgers}}]{Charnley1997}
{Charnley}, S.~B., {Tielens}, A.~G.~G.~M., \& {Rodgers}, S.~D. 1997, \apjl,
  482, L203

\bibitem[{{Clayton}(1993)}]{Clayton1993}
{Clayton}, R.~N. 1993, Annual Review of Earth and Planetary Sciences, 21, 115

\bibitem[{{Favre} {et~al.}(2011){Favre}, {Despois}, {Brouillet}, {Baudry},
  {Combes}, {Gu{\'e}lin}, {Wootten}, \& {Wlodarczak}}]{Favre2011}
{Favre}, C., {Despois}, D., {Brouillet}, N., {et~al.} 2011, \aap, 532, A32

\bibitem[{{Fisher} {et~al.}(2007){Fisher}, {Paciga}, {Xu}, {Zhao}, {Moruzzi},
  \& {Lees}}]{Fisher2007}
{Fisher}, J., {Paciga}, G., {Xu}, L.-H., {et~al.} 2007, Journal of Molecular
  Spectroscopy, 245, 7

\bibitem[{{Gardner} {et~al.}(1989){Gardner}, {Whiteoak}, {Reynolds}, {Peters},
  \& {Kuiper}}]{Gardner1989}
{Gardner}, F.~F., {Whiteoak}, J.~B., {Reynolds}, J., {Peters}, W.~L., \&
  {Kuiper}, T.~B.~H. 1989, \mnras, 240, 35P

\bibitem[{{Garrod} {et~al.}(2007){Garrod}, {Wakelam}, \& {Herbst}}]{Garrod2007}
{Garrod}, R.~T., {Wakelam}, V., \& {Herbst}, E. 2007, \aap, 467, 1103

\bibitem[{{Gerry} {et~al.}(1976){Gerry}, {Lees}, \& {Winnewisser}}]{Gerry1976}
{Gerry}, M.~C.~L., {Lees}, R.~M., \& {Winnewisser}, G. 1976, Journal of
  Molecular Spectroscopy, 61, 231

\bibitem[{{Henkel} {et~al.}(1987){Henkel}, {Jacq}, {Mauersberger}, {Menten}, \&
  {Steppe}}]{Henkel1987}
{Henkel}, C., {Jacq}, T., {Mauersberger}, R., {Menten}, K.~M., \& {Steppe}, H.
  1987, \aap, 188, L1

\bibitem[{{Henkel} \& {Mauersberger}(1993)}]{Henkel1993}
{Henkel}, C. \& {Mauersberger}, R. 1993, \aap, 274, 730

\bibitem[{{Hirota} {et~al.}(2007){Hirota}, {Bushimata}, {Choi}, {Honma},
  {Imai}, {Iwadate}, {Jike}, {Kameno}, {Kameya}, {Kamohara}, {Kan-Ya},
  {Kawaguchi}, {Kijima}, {Kim}, {Kobayashi}, {Kuji}, {Kurayama}, {Manabe},
  {Maruyama}, {Matsui}, {Matsumoto}, {Miyaji}, {Nagayama}, {Nakagawa},
  {Nakamura}, {Oh}, {Omodaka}, {Oyama}, {Sakai}, {Sasao}, {Sato}, {Sato},
  {Shibata}, {Shintani}, {Tamura}, {Tsushima}, \& {Yamashita}}]{Hirota2007}
{Hirota}, T., {Bushimata}, T., {Choi}, Y.~K., {et~al.} 2007, \pasj, 59, 897

\bibitem[{{Hoshino} {et~al.}(1996){Hoshino}, {Ohishi}, {Akabane}, {Ukai},
  {Tsunekawa}, \& {Takagi}}]{Hoshino1996}
{Hoshino}, Y., {Ohishi}, M., {Akabane}, K., {et~al.} 1996, \apjs, 104, 317

\bibitem[{{Hughes} {et~al.}(1951){Hughes}, {Good}, \& {Coles}}]{Hughes1951}
{Hughes}, R.~H., {Good}, W.~E., \& {Coles}, D.~K. 1951, Physical Review, 84,
  418

\bibitem[{{Ikeda} {et~al.}(1998){Ikeda}, {Duan}, {Tsunekawa}, \&
  {Takagi}}]{Ikeda1998}
{Ikeda}, M., {Duan}, Y.-B., {Tsunekawa}, S., \& {Takagi}, K. 1998, \apjs, 117,
  249

\bibitem[{{Larson}(1976)}]{Larson1976}
{Larson}, R.~B. 1976, \mnras, 176, 31

\bibitem[{{Mart{\'\i}n} {et~al.}(2019){Mart{\'\i}n}, {Muller}, {Henkel},
  {Meier}, {Aladro}, {Sakamoto}, \& {van der Werf}}]{Martin2019}
{Mart{\'\i}n}, S., {Muller}, S., {Henkel}, C., {et~al.} 2019, \aap, 624, A125

\bibitem[{{Mauersberger} {et~al.}(1988){Mauersberger}, {Henkel}, {Jacq}, \&
  {Walmsley}}]{Mauersberger1988}
{Mauersberger}, R., {Henkel}, C., {Jacq}, T., \& {Walmsley}, C.~M. 1988, \aap,
  194, L1

\bibitem[{{Melnick} {et~al.}(2010){Melnick}, {Tolls}, {Neufeld}, {Bergin},
  {Phillips}, {Wang}, {Crockett}, {Bell}, {Blake}, {Cabrit}, {Caux},
  {Ceccarelli}, {Cernicharo}, {Comito}, {Daniel}, {Dubernet}, {Emprechtinger},
  {Encrenaz}, {Falgarone}, {Gerin}, {Giesen}, {Goicoechea}, {Goldsmith},
  {Herbst}, {Joblin}, {Johnstone}, {Langer}, {Latter}, {Lis}, {Lord}, {Maret},
  {Martin}, {Menten}, {Morris}, {M{\"u}ller}, {Murphy}, {Ossenkopf}, {Pagani},
  {Pearson}, {P{\'e}rault}, {Plume}, {Qin}, {Salez}, {Schilke}, {Schlemmer},
  {Stutzki}, {Trappe}, {van der Tak}, {Vastel}, {Yorke}, {Yu}, \&
  {Zmuidzinas}}]{Melnick2010}
{Melnick}, G.~J., {Tolls}, V., {Neufeld}, D.~A., {et~al.} 2010, \aap, 521, L27

\bibitem[{{Menten} {et~al.}(2007){Menten}, {Reid}, {Forbrich}, \&
  {Brunthaler}}]{Menten2007}
{Menten}, K.~M., {Reid}, M.~J., {Forbrich}, J., \& {Brunthaler}, A. 2007, \aap,
  474, 515

\bibitem[{{Milam} {et~al.}(2005){Milam}, {Savage}, {Brewster}, {Ziurys}, \&
  {Wyckoff}}]{Milam2005}
{Milam}, S.~N., {Savage}, C., {Brewster}, M.~A., {Ziurys}, L.~M., \& {Wyckoff},
  S. 2005, \apj, 634, 1126

\bibitem[{{M{\"u}ller} {et~al.}(2024){M{\"u}ller}, {Ilyushin}, {Belloche},
  {Lewen}, \& {Schlemmer}}]{Muller2024}
{M{\"u}ller}, H. S.~P., {Ilyushin}, V.~V., {Belloche}, A., {Lewen}, F., \&
  {Schlemmer}, S. 2024, \aap, 688, A201

\bibitem[{{M{\"u}ller} {et~al.}(2001){M{\"u}ller}, {Thorwirth}, {Roth}, \&
  {Winnewisser}}]{Muller2001}
{M{\"u}ller}, H.~S.~P., {Thorwirth}, S., {Roth}, D.~A., \& {Winnewisser}, G.
  2001, \aap, 370, L49

\bibitem[{{Neill} {et~al.}(2013){Neill}, {Wang}, {Bergin}, {Crockett}, {Favre},
  {Plume}, \& {Melnick}}]{Neill2013}
{Neill}, J.~L., {Wang}, S., {Bergin}, E.~A., {et~al.} 2013, \apj, 770, 142

\bibitem[{{Pagani} {et~al.}(2017){Pagani}, {Favre}, {Goldsmith}, {Bergin},
  {Snell}, \& {Melnick}}]{Pagani2017}
{Pagani}, L., {Favre}, C., {Goldsmith}, P.~F., {et~al.} 2017, \aap, 604, A32

\bibitem[{{Peng} {et~al.}(2012){Peng}, {Despois}, {Brouillet}, {Parise}, \&
  {Baudry}}]{Peng2012}
{Peng}, T.~C., {Despois}, D., {Brouillet}, N., {Parise}, B., \& {Baudry}, A.
  2012, \aap, 543, A152

\bibitem[{{Penzias}(1981)}]{Penzias1981}
{Penzias}, A.~A. 1981, \apj, 249, 518

\bibitem[{{Persson} {et~al.}(2007){Persson}, {Olofsson}, {Koning}, {Bergman},
  {Bernath}, {Black}, {Frisk}, {Geppert}, {Hasegawa}, {Hjalmarson}, {Kwok},
  {Larsson}, {Lecacheux}, {Nummelin}, {Olberg}, {Sandqvist}, \&
  {Wirstr{\"o}m}}]{Persson2007}
{Persson}, C.~M., {Olofsson}, A.~O.~H., {Koning}, N., {et~al.} 2007, \aap, 476,
  807

\bibitem[{{Pickett} {et~al.}(1998){Pickett}, {Poynter}, {Cohen}, {Delitsky},
  {Pearson}, \& {M{\"u}ller}}]{Pickett1998}
{Pickett}, H.~M., {Poynter}, R.~L., {Cohen}, E.~A., {et~al.} 1998, \jqsrt, 60,
  883

\bibitem[{{Plume} {et~al.}(2012){Plume}, {Bergin}, {Phillips}, {Lis}, {Wang},
  {Crockett}, {Caux}, {Comito}, {Goldsmith}, \& {Schilke}}]{Plume2012}
{Plume}, R., {Bergin}, E.~A., {Phillips}, T.~G., {et~al.} 2012, \apj, 744, 28

\bibitem[{{Predoi-Cross} {et~al.}(1997){Predoi-Cross}, {Lees}, {Lichau},
  {Winnewisser}, \& {Drummond}}]{Predoi1997}
{Predoi-Cross}, A., {Lees}, R.~M., {Lichau}, H., {Winnewisser}, M., \&
  {Drummond}, J.~R. 1997, International Journal of Infrared and Millimeter
  Waves, 18, 2047

\bibitem[{{Romano}(2022)}]{Romano2022}
{Romano}, D. 2022, \aapr, 30, 7

\bibitem[{{Sch{\"o}ier} {et~al.}(2005){Sch{\"o}ier}, {van der Tak}, {van
  Dishoeck}, \& {Black}}]{Schoier2005}
{Sch{\"o}ier}, F.~L., {van der Tak}, F.~F.~S., {van Dishoeck}, E.~F., \&
  {Black}, J.~H. 2005, \aap, 432, 369

\bibitem[{{Shaver} {et~al.}(1983){Shaver}, {McGee}, {Newton}, {Danks}, \&
  {Pottasch}}]{Shaver1983}
{Shaver}, P.~A., {McGee}, R.~X., {Newton}, L.~M., {Danks}, A.~C., \&
  {Pottasch}, S.~R. 1983, \mnras, 204, 53

\bibitem[{{Soma} {et~al.}(2015){Soma}, {Sakai}, {Watanabe}, \&
  {Yamamoto}}]{Soma2015}
{Soma}, T., {Sakai}, N., {Watanabe}, Y., \& {Yamamoto}, S. 2015, \apj, 802, 74

\bibitem[{{Tamanai} {et~al.}(2025){Tamanai}, {Oyama}, {Watanabe}, {Sakai},
  {Nakatani}, {Zeng}, {Kleiner}, \& {Sakai}}]{Tamanai2025}
{Tamanai}, A., {Oyama}, T., {Watanabe}, Y., {et~al.} 2025, \apj, 980, 110

\bibitem[{{Tercero} {et~al.}(2010){Tercero}, {Cernicharo}, {Pardo}, \&
  {Goicoechea}}]{Tercero2010}
{Tercero}, B., {Cernicharo}, J., {Pardo}, J.~R., \& {Goicoechea}, J.~R. 2010,
  \aap, 517, A96

\bibitem[{{Tercero} {et~al.}(2018){Tercero}, {Cuadrado}, {L{\'o}pez},
  {Brouillet}, {Despois}, \& {Cernicharo}}]{Tercero2018}
{Tercero}, B., {Cuadrado}, S., {L{\'o}pez}, A., {et~al.} 2018, \aap, 620, L6

\bibitem[{{Tercero} {et~al.}(2012){Tercero}, {Margul{\`e}s}, {Carvajal},
  {Motiyenko}, {Huet}, {Alekseev}, {Kleiner}, {Guillemin}, {M{\o}llendal}, \&
  {Cernicharo}}]{Tercero2012}
{Tercero}, B., {Margul{\`e}s}, L., {Carvajal}, M., {et~al.} 2012, \aap, 538,
  A119

\bibitem[{{Tielens} \& {Whittet}(1997)}]{Tielens1997}
{Tielens}, A.~G.~G.~M. \& {Whittet}, D.~C.~B. 1997, in IAU Symposium, Vol. 178,
  IAU Symposium, ed. E.~F. {van Dishoeck}, 45

\bibitem[{{van Dishoeck} {et~al.}(1995){van Dishoeck}, {Blake}, {Jansen}, \&
  {Groesbeck}}]{Dishoeck1995}
{van Dishoeck}, E.~F., {Blake}, G.~A., {Jansen}, D.~J., \& {Groesbeck}, T.~D.
  1995, \apj, 447, 760

\bibitem[{{van Gelder} {et~al.}(2022){van Gelder}, {Jaspers}, {Nazari},
  {Ahmadi}, {van Dishoeck}, {Beltr{\'a}n}, {Fuller}, {S{\'a}nchez-Monge}, \&
  {Schilke}}]{Gelder2022}
{van Gelder}, M.~L., {Jaspers}, J., {Nazari}, P., {et~al.} 2022, \aap, 667,
  A136

\bibitem[{{Watanabe} \& {Kouchi}(2002)}]{Watanabe2002}
{Watanabe}, N. \& {Kouchi}, A. 2002, \apjl, 571, L173

\bibitem[{{Watanabe} {et~al.}(2021){Watanabe}, {Chiba}, {Sakai}, {Tamanai},
  {Suzuki}, \& {Sakai}}]{Watanabe2021}
{Watanabe}, Y., {Chiba}, Y., {Sakai}, T., {et~al.} 2021, \pasj, 73, 372

\bibitem[{{Watanabe} {et~al.}(2014){Watanabe}, {Sakai}, {Sorai}, \&
  {Yamamoto}}]{Watanabe2014}
{Watanabe}, Y., {Sakai}, N., {Sorai}, K., \& {Yamamoto}, S. 2014, \apj, 788, 4

\bibitem[{{Wilson}(1999)}]{Wilson1999}
{Wilson}, T.~L. 1999, Reports on Progress in Physics, 62, 143

\bibitem[{{Wouterloot} {et~al.}(2008){Wouterloot}, {Henkel}, {Brand}, \&
  {Davis}}]{Wouterloot2008}
{Wouterloot}, J.~G.~A., {Henkel}, C., {Brand}, J., \& {Davis}, G.~R. 2008,
  \aap, 487, 237

\bibitem[{{Wright} {et~al.}(1996){Wright}, {Plambeck}, \&
  {Wilner}}]{Wright1996}
{Wright}, M.~C.~H., {Plambeck}, R.~L., \& {Wilner}, D.~J. 1996, \apj, 469, 216

\bibitem[{{Yang} {et~al.}(2021){Yang}, {Sakai}, {Zhang}, {Murillo}, {Zhang},
  {Higuchi}, {Zeng}, {L{\'o}pez-Sepulcre}, {Yamamoto}, {Lefloch}, {Bouvier},
  {Ceccarelli}, {Hirota}, {Imai}, {Oya}, {Sakai}, \& {Watanabe}}]{Yang2021}
{Yang}, Y.-L., {Sakai}, N., {Zhang}, Y., {et~al.} 2021, \apj, 910, 20

\bibitem[{{Yurimoto} \& {Kuramoto}(2004)}]{Yurimoto2004}
{Yurimoto}, H. \& {Kuramoto}, K. 2004, Science, 305, 1763

\bibitem[{{Zhang} {et~al.}(2020){Zhang}, {Liu}, {Yan}, {Yu}, {Liu}, {Zheng},
  {Romano}, {Zhang}, {Wang}, {Chen}, {Wang}, {Zhang}, {Lu}, {Chen}, {Zou},
  {Yang}, {Wen}, \& {Lu}}]{Zhang2020}
{Zhang}, J.~S., {Liu}, W., {Yan}, Y.~T., {et~al.} 2020, \apjs, 249, 6

\bibitem[{{Zou} {et~al.}(2023){Zou}, {Zhang}, {Henkel}, {Romano}, {Liu},
  {Zheng}, {Yan}, {Chen}, {Wang}, \& {Zhao}}]{Zou2023}
{Zou}, Y.~P., {Zhang}, J.~S., {Henkel}, C., {et~al.} 2023, \apjs, 268, 56

\end{thebibliography}

\end{document}